\documentstyle[sprocl,epsfig]{article}

\def\Journal#1#2#3#4{{#1} {\bf #2}, #3 (#4)}

\def\PRD{{\em Phys. Rev.} D}

\def\be{\begin{equation}}
\def\ee{\end{equation}}
\def\bea{\begin{eqnarray}}
\def\eea{\end{eqnarray}}

\begin{document}

\begin{flushright}
BROWN-HET-1178 TA-570
\end{flushright}
\vskip 1.0cm

\title{HIGH ENERGY FLUXES FROM A NON-SCALING COSMIC STRING 
NETWORK}

\author{ U.F. WICHOSKI, R.H. BRANDENBERGER }

\address{Department of Physics, Brown University, \\ 
Box 1843, Providence, RI 02912, USA }

\author{ J.H. MACGIBBON }

\address{Code SN3, NASA Johnson Space Center, \\ 
Houston, TX 77058, USA } 


\maketitle\abstracts{
Topological defects, particularly cosmic strings, can 
provide a mechanism to produce particles with energies of the order 
$10^{21}$ eV and higher. Here, we report on order of 
magnitude calculations of fluxes from a cosmic string network 
which evolves according to a new scenario according to which 
the main channel 
for energy loss is the particle production rather than gravitational 
radiation. We compare the predicted fluxes for protons (anti-protons) and 
neutrinos (anti-neutrinos) with observations of extremely 
high energy cosmic rays.} 
  
\section{Introduction}

Cosmic strings~\cite{cs} 
are linear topological defects predicted to arise in many 
particle physics models during a symmetry breaking 
phase transition in the early Universe. 
Cosmic strings can be relevant 
for structure formation,\cite{bran2} but they can also be 
important as a source of extremely high energy cosmic rays.\cite{bha}  

Recently, cosmic rays events with energies above $10^{20}$ eV 
were detected by various experiments.\cite{exp} 
The origin of these events is unknown to date. 
There are two main scenarios.  
The first is astrophysical 
and is based on the idea that charged particles are 
accelerated in shocks. 
Specifically, in the case of the 
extremely high energy cosmic rays, these shocks are most likely 
associated with active galactic nuclei (AGNs) and 
powerful radio galaxies. The major problems in the 
acceleration scenario, or `bottom-up' scenario, is that 
in the case of AGNs the energy 
gained by the particle is mostly lost in collisions with the 
medium within which the acceleration takes place. In the case 
of radio galaxies this is not of much concern 
although the distance at which these objects are located 
($> 100$ Mpc) constitutes a problem (see e.g. reference~\cite{bs} 
and refs. therein). 

The other possibility is that the the decay products of 
very massive particles produced in the early Universe 
are the source of the extremely high energy 
protons (anti-protons), gamma-rays and neutrinos (anti-neutrinos).
In this scenario, also known as the `top-down' scenario,\cite{bs} 
no acceleration 
is needed since these very massive particles, which are referred to 
as X particles, can be as heavy as $10^{16}$ GeV. 

In this paper, we report on an order of magnitude 
calculation of the fluxes of extremely high energy protons 
(anti-protons) and neutrinos (anti-neutrinos) in the scenario 
in which the particle production from a cosmic string network is 
maximal (VHS scenario.\cite{vhs}) 

\section{Standard Cosmic String and VHS Scenarios} 

In many particle physics models of matter, linear topological defects 
(cosmic strings) will be produced during a phase transition in the 
early Universe. Strings are topologically stable configurations in 
the core of which the superheavy Higgs and gauge particles which 
obtain a mass during the phase transition are trapped. 
Strings arise since the fields are uncorrelated in regions separated 
by more than the thermal correlation length $\xi$, which  
by causality at time $t$ must be smaller than the horizon $t$. Strings 
are characterized by the mass per unit length $\mu \simeq \eta^2$, 
where $\eta$ is the energy scale of the symmetry breaking.  Right 
after formation, the strings are in a random tangle
d configuration which subsequently tends to straighten itself out. 
Any nongravitational string decay corresponds to the emission and 
subsequent decay of X particles into jets of high energy  particles. 

The conformal stretching due to the expansion of the Universe alone, 
would lead to a Universe dominated by strings. One can show, however, 
that the network of long strings (strings with curvature radius larger 
than the Hubble radius) must steadily decay and achieve 
a ``scaling solution" in which 
all lengths scale with the Hubble radius and hence
\[
\rho_{\infty} = {{\nu \mu} \over {t^2}} \, ,
\]
where $\nu$ is the number of strings per Hubble volume. 
In the standard cosmic string scenario, the decay mechanism 
is provided by the (predominantly gravitational) decay of 
cosmic strings loops formed through the 
intersection (self-intersection) and reconnection of long 
cosmic strings.\cite{cs} 
In contrast, according to the VHS scenario,\cite{vhs} long cosmic 
strings release their energy directly into X particles.
 
\section{Particle Production}

The energy conservation equation for the network of long cosmic 
strings is 
\[
{\dot \rho_{\infty}} + 2 H \rho_{\infty} \, = 
\, - m_X {{d n_X} \over {dt}} \,
\]
where $H$ is the Hubble parameter, and 
$n_X$ and $m_X$ are the number density and mass, respectively, 
of the X-particles. 

The decay of the X-particles leads to the production 
of jets. We assume that the initial 
energy $m_J$ of all jets is the same. In this case, the decay of 
a single X-particle will lead to $m_X / m_J$ jets, and the number 
density of jets resulting from the energy release of long strings is
\be \label{jetdensity}
{{d n_J} \over {dt}} \, = {{\nu \mu} \over {m_J}} 
t^{-3} \, . 
\ee
Because of our ignorance of the structure of the jets at extremely 
high 
initial energies, we extrapolate the QCD fragmentation function of 
the jets into quarks and leptons   
(known, at least as a good approximation, up to a few TeV). 
This is, of course, the source 
of the largest uncertainty in the calculation of the fluxes. 
The distribution of energies $E$ of the primary decay products of 
the jet can be well approximated by a fragmentation function 
based on a simple $E^{1/2}$ multiplicity 
\be \label{primdecay}
{{d {N'}} \over {d x}} \, = \, {{15} \over {16}} x^{-3/2} 
(1 - x)^2 \, ,
\ee
where $x = E / m_J$ is the fraction of the jet energy which the decay 
product receives.

The initial jet particle decays into quarks and leptons on a time 
scale of $\alpha m_J^{-1}$, where $\alpha$ is coupling constant 
associated with the physics at an energy scale of $m_J$. The quarks 
then hadronize on a strong interaction time scale. Most of the 
energy (about 97\%) goes into pions, the remainder into baryons. 
The neutral pions decay into two photons, the charged pions decay 
by emitting neutrinos. Note that the contribution of the primary 
leptons to the total flux of leptons is negligible. 
Integrating (\ref{primdecay}) from $x$ to $1$ 
with the invariant measure $dx / x$, we obtain the distribution 
of the 
energies of the products of two body decays of the primary particles 
\be \label{secdecay}
{{d N} \over {d x}} \, = \, {{15} \over {16}} \bigl[{{16} 
\over 3} - 2 x^{1/2} - 4 x^{-1/2} + {2 \over 3} x^{-3/2} \bigr]\, .
\ee
Equation (\ref{secdecay}) applies to the spectrum of neutrinos 
produced in the jet, whereas equation (\ref{primdecay}) 
applies to the primary decay products such as protons 
(anti-protons). Since only about 3\% of the energy of the jet 
goes into primary protons (anti-protons), the distribution 
of these particles is given by (\ref{primdecay}) multiplied 
by the factor 0.03.

The expressions (\ref{jetdensity}) and (\ref{primdecay}) or 
(\ref{secdecay}) for the number density of jets and for the 
energy distribution of the jet decay products can be 
combined to obtain 
the expected fractional flux $F(E)$ of high energy neutrinos 
and cosmic ray protons of energy $E$ produced in the VHS cosmic 
string scenario. The general formula is 
\be \label{flux}
F(E) \, = \, \int_{t_c}^{t_{isd}} dt' e^{-t/t_{c}} {{d n_J} \over {d t'}} 
(z(t') + 1)^{-3} {{d N} \over {d E'}} (z(t') + 1) \, ,
\ee
where $t_c$ is the earliest emission time for a particle 
with present energy 
$E$, and $t_{isd}$ is the latest time that 
the emission from the cosmic string network can be considered to 
have isotropized by the present time.\cite{ujr} 
The most important propagation effects for protons (anti-protons) 
of extremely high energies are:\cite{bs,bb} 
{\it i-)} pair production and 
{\it ii-)} photoproduction of pions (GZK~\cite{gzk} cutoff at 
energies above $10^{11}$ GeV drastically limits the distance to 
the source). In the case of neutrinos,\cite{bs,yos} 
the leading energy loss effect is the interaction 
with the (presently) $1.9$ K cosmic neutrino background. 
The propagation effects determine the limits 
in the integral (\ref{flux}). 

\section{Diffuse fluxes}

In order to calculate the diffuse flux of protons and neutrinos 
we have to take into account  that in the VHS scenario 
cosmic string loops 
collapse almost immediately after their formation. Therefore, the 
X particles are produced along the string. Because the inter-string 
distance grows as the Universe expands, eventually this distance is 
bigger than the attenuation length for the propagation of the 
particles. For the propagation of neutrinos this effect is small 
because neutrinos interact only weakly and their 
attenuation length is comparable to the horizon. On the other hand, 
protons at extremely high energies have an attenuation length 
smaller than $\sim 100$ Mpc. Therefore, the flux is 
exponentially suppressed as the inter-string distance 
increases beyond this value.\cite{be} 

\begin{figure}
\psfig{figure=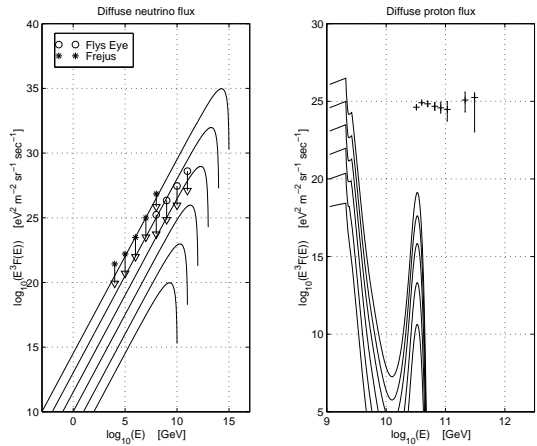,height=2.5in} 
\caption[...]{Diffuse neutrino flux (left) and diffuse proton 
flux (right) in the VHS cosmic string scenario for 
various values of $G \mu$ (from top to bottom, 
$G \mu = 10^{-6}$, $10^{-8}, 10^{-10}, 10^{-12}, 10^{-14}, 10^{-16}$). 
Points with arrows represent upper limits on the diffuse neutrino 
flux from the Fr\'ejus~\cite{frejus} and the 
Fly's Eye~\cite{eye} experiments. Points with error bars correspond 
to the combined cosmic ray data from the Fly's Eye and AGASA 
experiments.\cite{exp}}
\end{figure}

Fig.(1) shows the fluxes from an order of magnitude calculation of
extremely high energy protons and neutrinos 
originating from the decay of X particles produced by cosmic string 
decay in the VHS scenario. Due the fact 
that the inter-string distance is much bigger than the attenuation 
length for protons, the proton flux is suppressed. The neutrino flux 
is shown for various values of $G \mu$ taking $m_J = \eta$ as the 
initial jet energy. 

\section{Conclusion}

The order of magnitude calculations we have performed here show that 
if a cosmic string network evolves as described by the VHS scenario, 
the flux of extremely high cosmic rays can be used to constrain the 
value of $G \mu$ to $G \mu < 10^{-10}$ (see Fig. 1). The predicted 
flux is dominated by neutrinos, and thus if the observed events are 
due to strings, they cannot have a proton or anti-proton 
as a primary.   

\section*{Acknowledgments}
This work has been supported (at Brown) in part by the US Department 
of Energy under contract DE-FG0291ER40688, Task A, and was performed 
while JHM held a NRC-NASA/JSC Senior Research Associateship. 
We would like to thank V. Berezinsky for his comments and suggestions. 
UFW is 
grateful to A. Mour\~ao and J.D. de Deus for the warm reception at 
CENTRA and to IST-CENTRA for financial support.

\end{document}